# A Novel Analysis Framework for Microstructural Characterization of Ferroelectric Hafnia: Experimental Validation and Application

Y. Park[1†], J. Jang[2†], H. Lee[2], K. Kim[1], K. Jung[1], Y. Lee[1], J. Lee[2], E. Yang[2], S. Jo[1], S. Yoo[1], H. J. Lee[3], D. Kim[1], D.-H. Choe[1]*, S. G. Nam[1]*

[1]Device Research Center (DRC), Samsung Advanced Institute of Technology (SAIT), 16678, Republic of Korea

[2]Material Research Center (MRC), SAIT, 16678, Republic of Korea

[3]School of Electrical and Computer Engineering, Georgia Institute of Technology, Atlanta GA 30332, USA

*Email: sg1118nam@samsung.com; dukhyun.choe@samsung.com †: Authors with equal contribution

***Abstract***—Herein, we present a novel analysis framework for grain size profile of ferroelectric hafnia to tackle critical shortcomings inherent in the current microstructural analysis. We vastly enhanced visibility of grains with ion beam treatment and performed accurate grain segmentation using deep neural network (DNN). By leveraging our new method, we discovered unexpected discrepancies that contradict previous results, such as deposition temperature ($T_{dep}$) and post-metallization annealing (PMA) dependence of grain size statistics, prompting us to reassess earlier interpretations. Combining microstructural analysis with electrical tests, we found that grain size reduction had both positive and negative outcomes: it caused significant diminishing of die-to-die variation (~68 % decrease in standard deviation) in coercive field ($E_c$), while triggering an upsurge in leakage current. These uncovered results signify robustness of our method in characterization of ferroelectric hafnia for in-depth examination of both device variability and reliability.

## I. Introduction

The ferroelectric hafnia exhibits an enormous potential to enable forthcoming ferroelectric memories with its unique properties, such as non-volatility and fast switching speed. In ferroelectric hafnia, the formation of grains is inherent due to its polycrystalline nature, and the grains are closely related to device performance in terms of ferroelectric properties and inter-cell variability of devices. A number of recent studies have investigated grain size statistics of Zr-doped Hafnium oxide (HZO) and their effects on ferroelectric properties [1-8], though they showed inconsistent results. For example, the mean grain sizes determined from ALD-grown ferroelectric HZO range from only a few nanometers [1,2] to several tens of [3-5], or even nearly a hundred nanometers [6]. Besides, some studies show discrete relationship between grain size and ferroelectric properties, such as inversely proportional relationship between grain size and remnant polarization ($P_r$) [1,3,5], while the others did not demonstrate such well-defined relationships [2,4].

Such inconsistency may be due to inaccurate grain size determination caused by several detrimental factors in widely-used analysis approaches. For example, indistinguishable grain boundaries in scanning electron microscopy (SEM) images by recent studies may worsen the accuracy of grain size statistics [2,7]. Moreover, conventionally adopted watershed algorithms, which have been implemented via the software named "Gwyddion" [8] for grain segmentations, have serious over-segmentation issues [9] that may cause underestimation of grain size statistics. To resolve such issues, we propose a novel method for obtaining accurate grain size statistics, described in **Fig. 1**. We performed ion beam treatment to the surface of HZO after metal layer etching to clean out surface residues thus enhancing grain boundary visibility. After grains were clearly visible, they were effectively segmented for extracting grain size statistics using DNN-based program. Once we obtained accurate grain size statistics using the method, we utilized it to establish a link with electrical test results for profound understanding of its effect on ferroelectric properties that previous studies might have overlooked.

## II. Methodology Development

### A. Sample Preparation

To validate our approach, we prepared HZO thin films with different thicknesses ($t$), $T_{PMA}$, PMA time, and $T_{dep}$ on two different stack structures: (from top) metal- ferroelectric- metal (MFM) and metal- ferroelectric- insulator- semiconductor (MFIS), since these process conditions are known to vary grain size of HZO [1-5]. The cross-section structural images of MFM and MFIS capacitors are shown in **Fig. 2**. SEM images with direct comparison of pre- and post-ion beam treatment are shown in **Fig. 3**. It is evident that, by applying ion beam treatment, surface residues created after top electrode etching are effectively removed and grain boundaries are clearly observed. To confirm that our observation is made on real polycrystalline grains of ferroelectric HZO, we made comparison between plan-view transmission electron microscopy (TEM) images and SEM images of HZO after ion beam treatment, as shown in **Fig. 4**. The direct comparison made in **Fig. 4** corroborates that grains revealed by SEM after FIB treatment are undoubtedly grains of HZO thin films.

### B. Deep Learning Processes

Though grains are clearly visible after FIB treatment, the contrast of images itself is not quantitative. Therefore, to perform effective grain segmentation in various contrast conditions, U-net, which is a DNN that is most widely known in image segmentation [10], is utilized. For the training of U-net, training data was generated by experts' works, which are images of grain segmentation performed with experts' hand-drawing. **Fig. 5** shows the training processes and results of a coupled U-net models using a SEM image, an inference (segmentation results), and an expert's work. The major difference between the coupled U-net models in this study is that there is a 2$^{nd}$ model, which is trained using the 1$^{st}$ result and labeled data, and is iteratively inferred to generate the image segmentation. We added 2nd model since the inference result of the 1st model (**Fig. 5(b)**, **N = 1**) could not find weak boundaries with low contrast signal; utilizing the 2nd model effectively compensated for this weakness and enabled finding the correct boundaries by repeatedly performing inference.

**Fig. 6(a)** shows SEM images before FIB treatment and segmentation results from conventional method, showing that no accurate grain size statistics (**Fig. 6(c)**) can be obtained without ion beam treatment. **Fig. 6(b)** shows segmentation results of watershed algorithm and DNN-based method, along with expert's work, after ion beam treatment. The statistical results in **Fig. 6(c)** show that, even though the image shows clear grain boundaries after ion beam treatment, watershed algorithms exhibit over-segmentation issues that causes underestimation of mean ($\mu$) and standard deviation ($\sigma$). On the contrary, DNN-based method effectively recognizes grains and the segmentation results and the corresponding statistics are almost identical to those of an expert's work.

## III. Results and Discussion

After validating our method, we employed our method to investigate the samples with various process conditions. First, we confirmed that the grain size statistics were not affected by either PMA time or PMA temperature ($T_{PMA}$), as illustrated in **Fig. 7**. The results contradict to $T_{PMA}$ dependence of grain size in the recent studies, in which both annealing time and temperature increase grain size [3,4,11]. Next, we examined the effect of $T_{dep}$ on grain size statistics. The boxplots and $\mu$ values for $T_{dep}$ dependence of grain size statistics of ferroelectric HZO with different $t$ (5 and 7 nm) in MFM stack are illustrated in **Fig. 8 (a)** and **(b),** respectively, along with corresponding SEM images shown in **Fig. 8 (c)**. These reveal that both $\mu$ and $\sigma$ of grain size significantly decrease as $T_{dep}$ increases, though no obvious $t$ dependence was observed. Unexpectedly, this $T_{dep}$ dependence in our study opposes to those from previous results, in which grain size increases as $T_{dep}$ increases [1,5]. Based on the trend in our study, it can be inferred that the grain size is intrinsically determined by the number of seed formed during deposition process (higher $T_{dep}$ provides more crystal seed) and is not varied during the afterward annealing processes, as illustrated in **Fig. 9**.

In addition, to investigate stack dependence of grain size, we acquired grain size statistics of HZO in MFM and MFIS stacks. The statistics of grain sizes each stack are illustrated as boxplots in **Fig. 10 (a)**, along with corresponding SEM images shown in **Fig. 10 (b)**. The grain size of HZO in MFIS stack is always larger than that of HZO in MFM stack, regardless of $T_{dep}$. The larger grain size in MFIS stack matches with the results from the recent study conducted with cross-section TEM images [6], in that grain size of $ZrO_2$ in MFIS stack is larger than that in MFM stack due to the local epitaxial templating effect from adjacent metallic layers. Therefore, it can be implied that our results align closely with those obtained using TEM, further affirming the validity of our method.

After we investigated grain size statistics of HZO stacks fabricated in various process conditions, we paid our attention to their electrical and ferroelectric properties. **Fig. 11** illustrates grain size dependence of DC leakage current density of HZO on MFM stacks taken at 2 V. It implies that smaller grain size causes more leakage. The grain boundaries usually act as leakage paths [12], and it can be understood that more grain boundaries with smaller grain sizes leads to more leakage current, considering formation of columnar grains along $t$.

P-V hysteresis curves of HZO with different $t$ and grain size in MFM stack are shown in **Fig. 12 (a)**. While grain size show significant drop from 56 nm to 32 ~ 33 nm, $P_r$ drops from 52 $\mu C/cm^2$ to 45 $\mu C/cm^2$ in case of $t$ = 7 nm, whereas it shows only a minor decrease in case of $t$ = 5 nm; Therefore, different from the previous studies, in which $P_r$ drops significantly as grain size increases [3,5], it can be concluded that grain size does not have significant impact on $P_r$, and a trivial decrease in $P_r$ might attribute to monoclinic phase formed due to higher $T_{dep}$. **Fig. 12 (b)** shows boxplots of $E_c$ taken from the hysteresis curves. Regardless of $t$, the $\mu$ of $E_c$ is show a notable increase, as grain size decreases. Here, an important finding is that the $\sigma$ of $E_c$ was remarkably reduced by 68 % and 42 % at $t$ = 5 nm and $t$ = 7 nm, respectively, as grain size decreased from 56 nm to 32 ~ 33 nm. These results imply that smaller grain size leads to increase in magnitude of $E_c$, and reduction in $\sigma$ of grain size (considering that, for grain size, reduction in $\mu$ causes reduction in $\sigma$ in general) leads to reduction in $\sigma$ of $E_c$.

## IV. Conclusion

By developing and harnessing a novel framework for microstructural characterization of ferroelectric hafnia, we have revealed unforeseen results that diverge from previous findings. We have presented converse relationship between $T_{dep}$ and grain size, as well as invariance of grain size under various PMA conditions. Moreover, we have also made key findings that grain size reduction significantly improved the die-to-die variability and a notable increase in $E_c$, while rapidly escalating leakage current. Advancement in current ferroelectric memory technology requires minimization of device variability, thus grain size reduction is vital; however, careful consideration of leakage current issue and mitigation strategy, such as doping [13] and layer insertion [14], should be accompanied to ensure development and commercialization of high-performance ferroelectric memory devices. Underscoring the need for advanced characterization methodologies, our method offers a means to identify and engineer the key characteristics of ferroelectric hafnia for future memory applications.

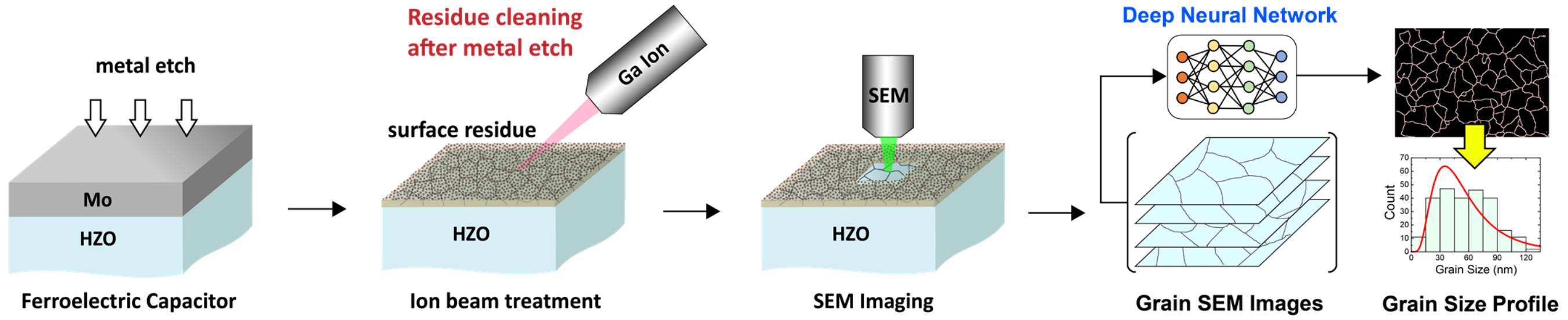

**Fig. 1.** A schematic illustration of complete process for obtaining grain size profiles of ferroelectric hafnia thin films. The main key processes are: surface cleaning process after metal etching using ion beam treatment, imaging polycrystalline grains with scanning electron microscopy (SEM) and automated grain segmentation process based on deep neural network (DNN).

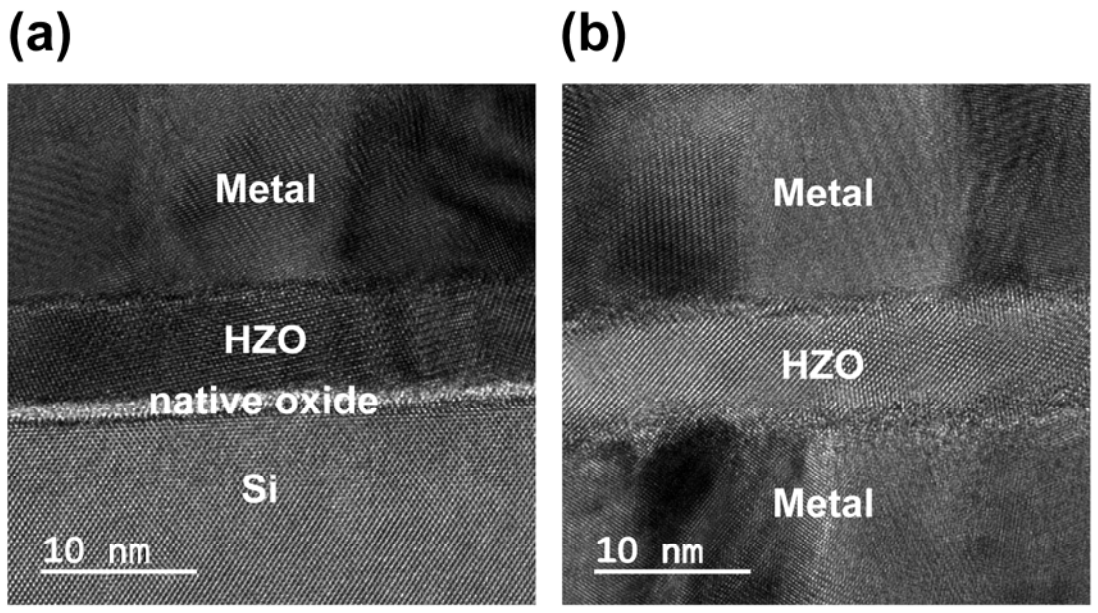

**Fig. 2.** Cross-section transmission electron microscopy (TEM) images of **(a)** MFIS and **(b)** MFM capacitors.

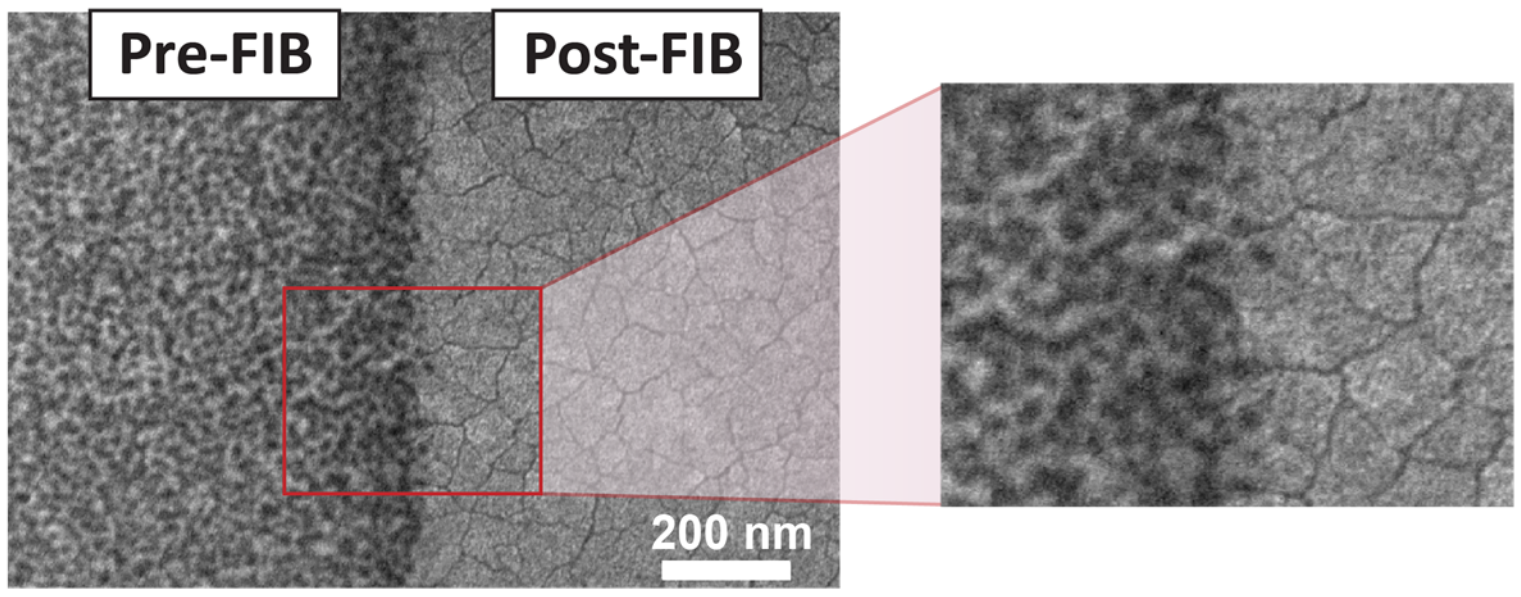

**Fig. 3.** A scanning electron microscopy (SEM) image with its partial magnification, of surface of HZO thin film showing the difference between surface morphology before and after FIB treatment.

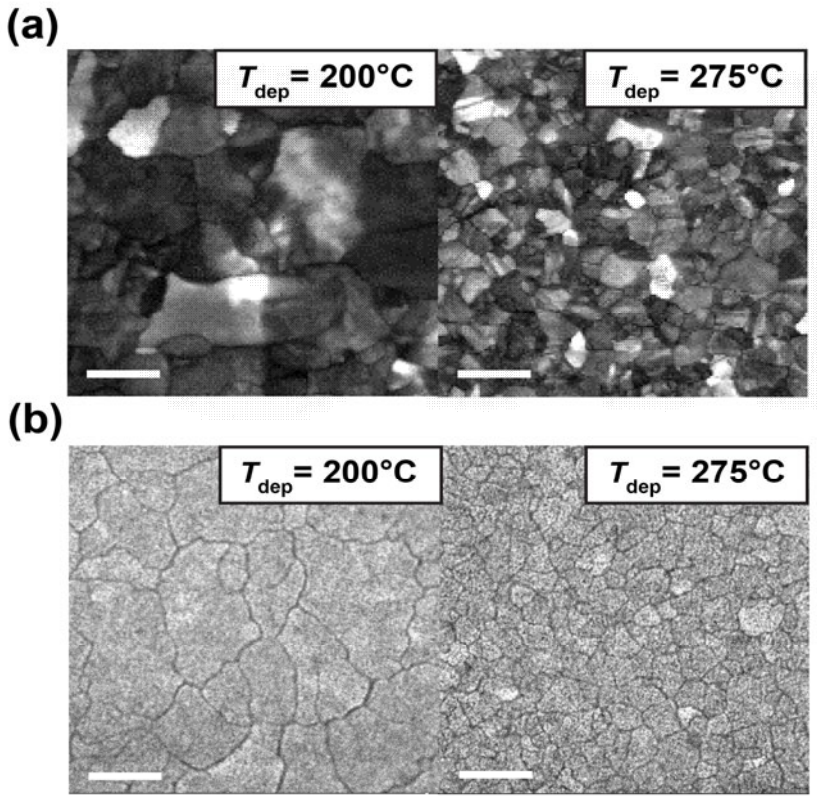

**Fig. 4.** Microstructural images of HZO thin films on MFIS stack. **(a)** Plan-view TEM images and **(b)** SEM images of HZO with different $T_{dep}$. White scale bar shows 200 nm.

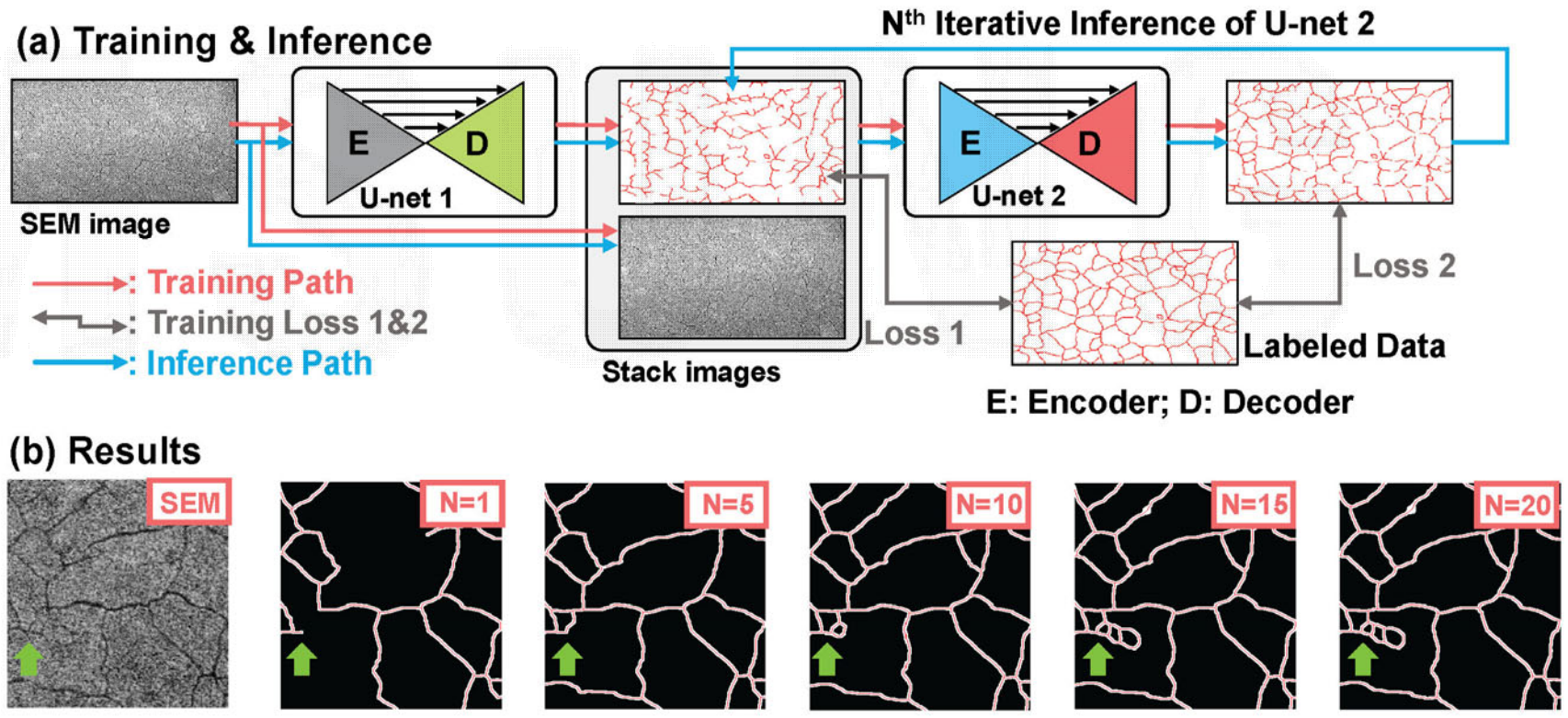

**Fig. 5** HZO Grain segmentation using deep neural network (DNN). **(a)** Schematic diagram of deep neural network (DNN) processes using coupled U-Net and **(b)** grain segmentation results with different iteration number (N).

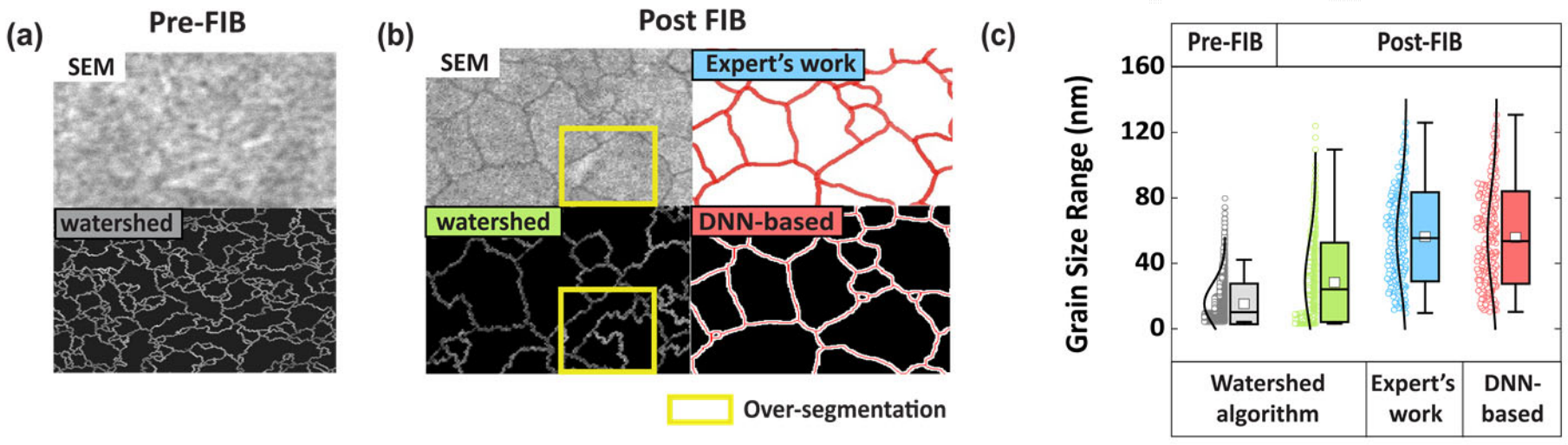

**Fig. 6** Comparison of conventional segmentation method and DNN-based method. **(a)** A SEM image of HZO thin film before FIB treatment and corresponding segmentation result using conventional (watershed) method **(b)** A SEM image after ion beam treatment and corresponding segmentation results from the expert's work, conventional method and DNN-based method. **(c)** Statistical results for comparison of different methods.

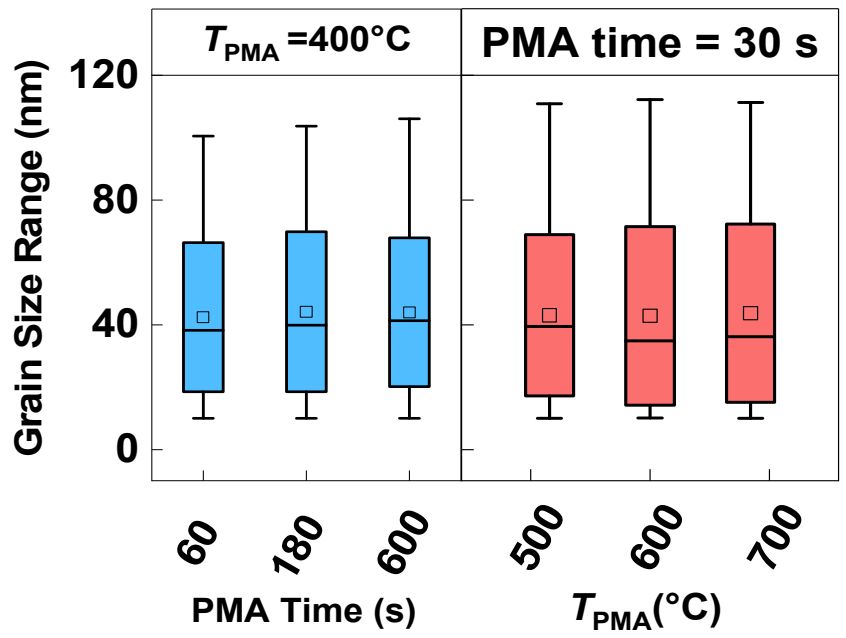

**Fig. 7.** Boxplots for PMA time and $T_{PMA}$ dependence of grain size statistics while fixing $T_{dep}$ at 250°C. Note invariance of statistical profiles under different PMA conditions.

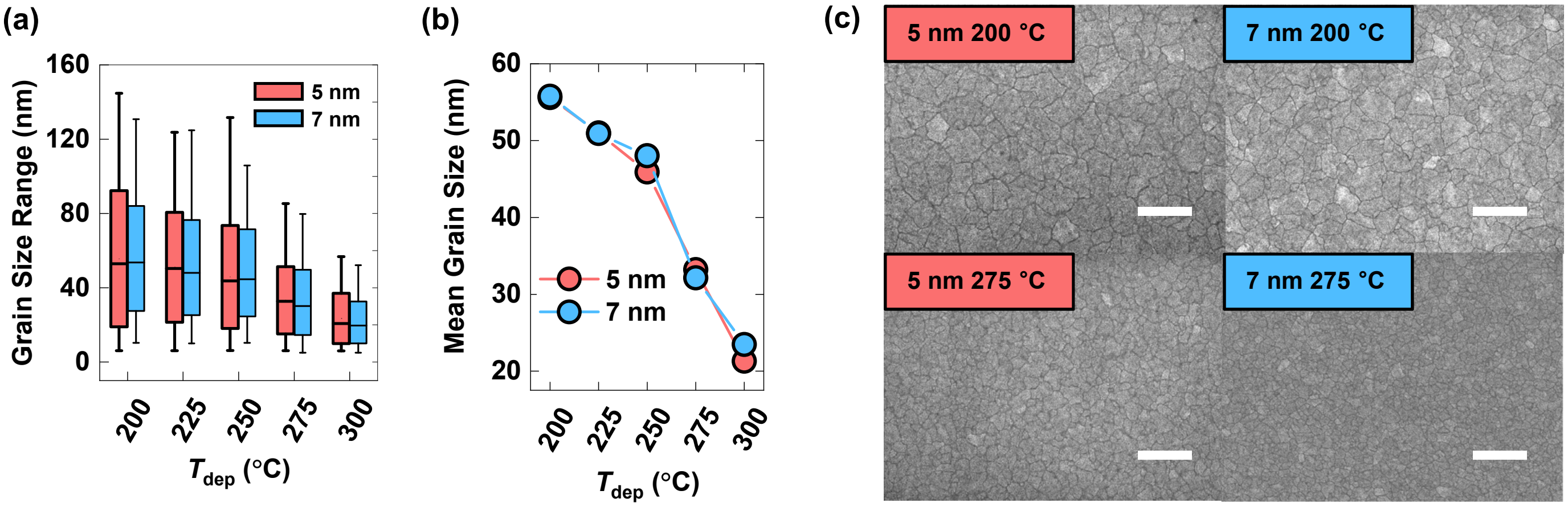

**Fig. 8.** $T_{dep}$ dependence of grain size of HZO thin films on MFM stack. **(a)** Boxplots for grain size statistics of HZO with different $T_{dep}$ and $t$. **(b)** $T_{dep}$ dependence of mean grain size. **(c)** SEM images of grains at $T_{dep}$ = 200 and 275 °C. White scale bar shows 200 nm.

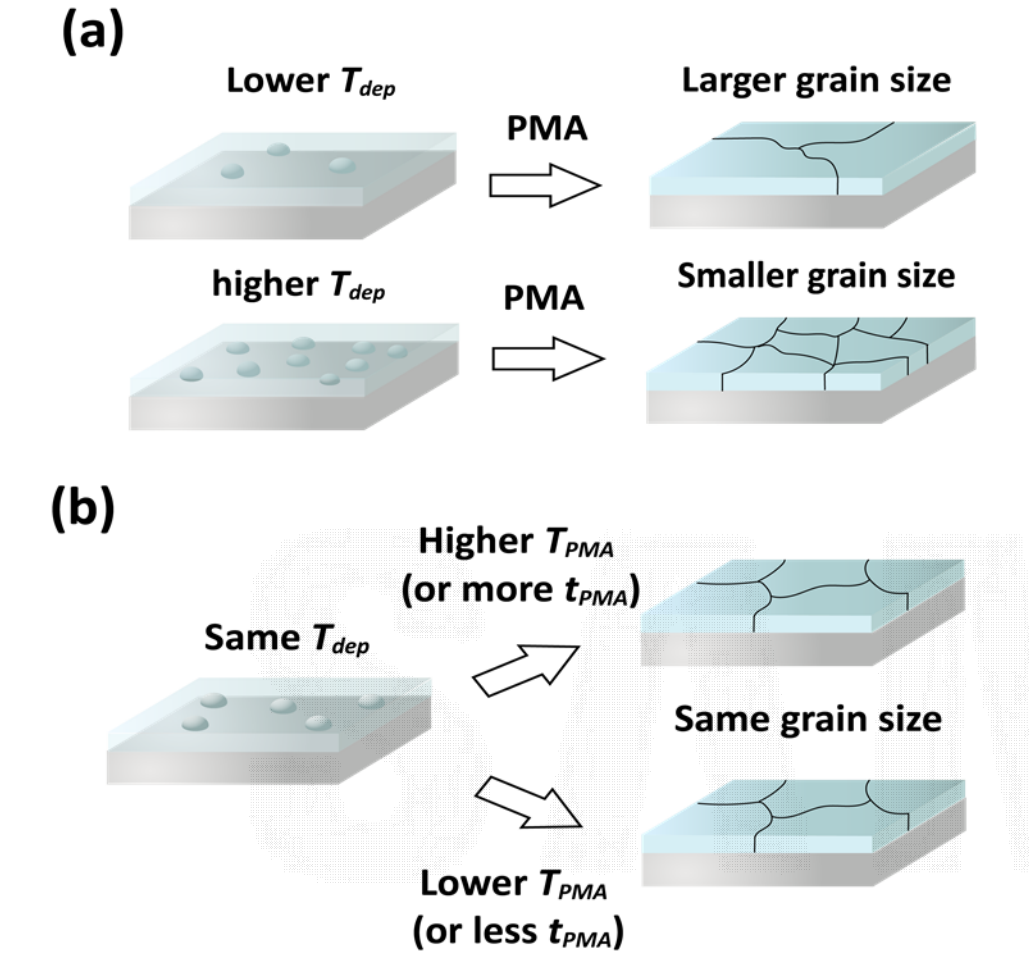

**Fig. 9.** Schematic illustration of grain growth mechanism of a ferroelectric HZO thin film. Grain formation **(a)** under different $T_{dep}$ and **(b)** under different PMA conditions, showing that grain size is intrinsically determined by the number of seed nucleation during deposition process.

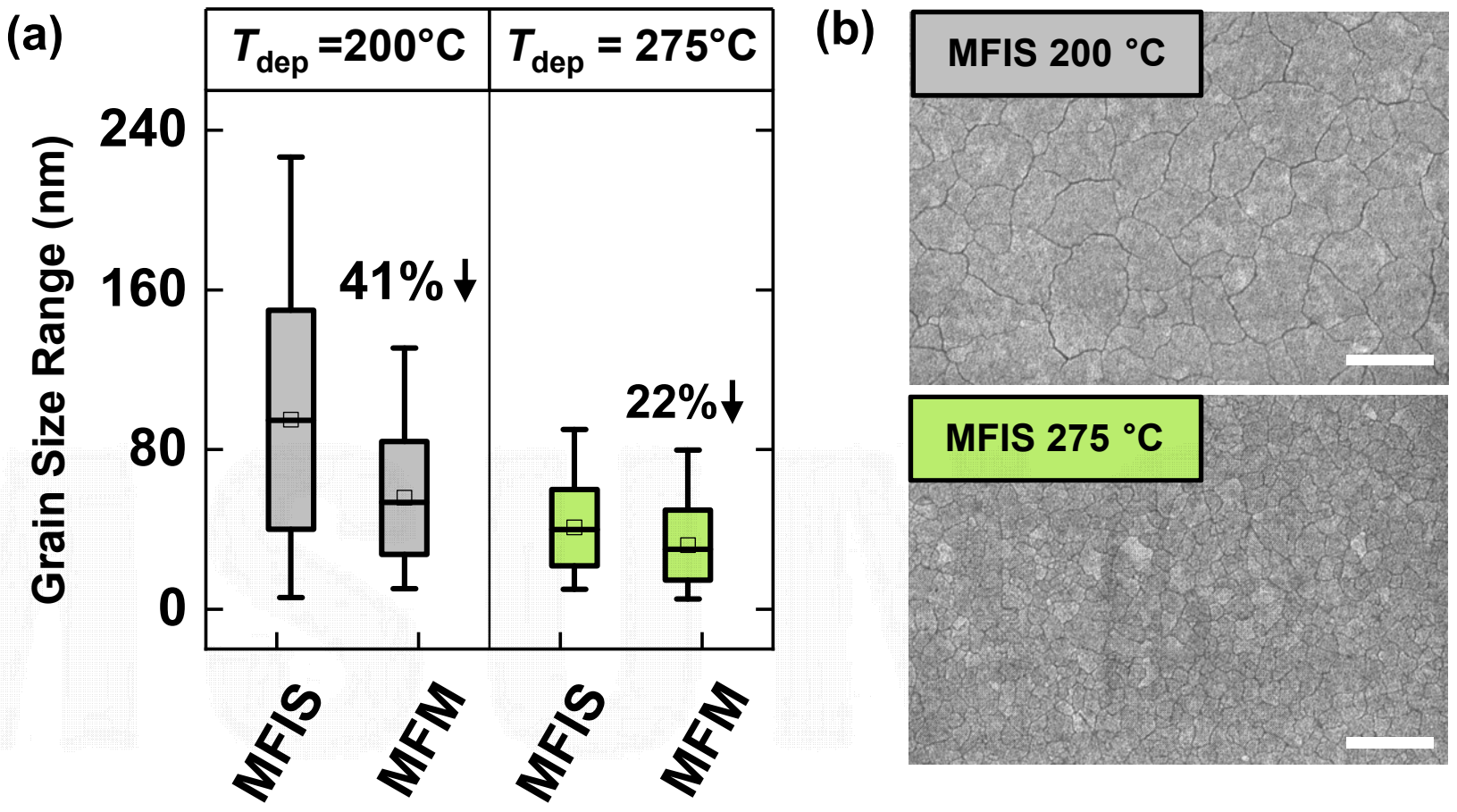

**Fig. 10.** Stack-dependence of grain size of HZO thin films. **(a)** Boxplots for grain size with different stack structures at $T_{dep}$ = 200 and 275 °C at $t$ = 7 nm. **(b)** Corresponding SEM images of grains in MFIS stack. White scale bar shows 200 nm.

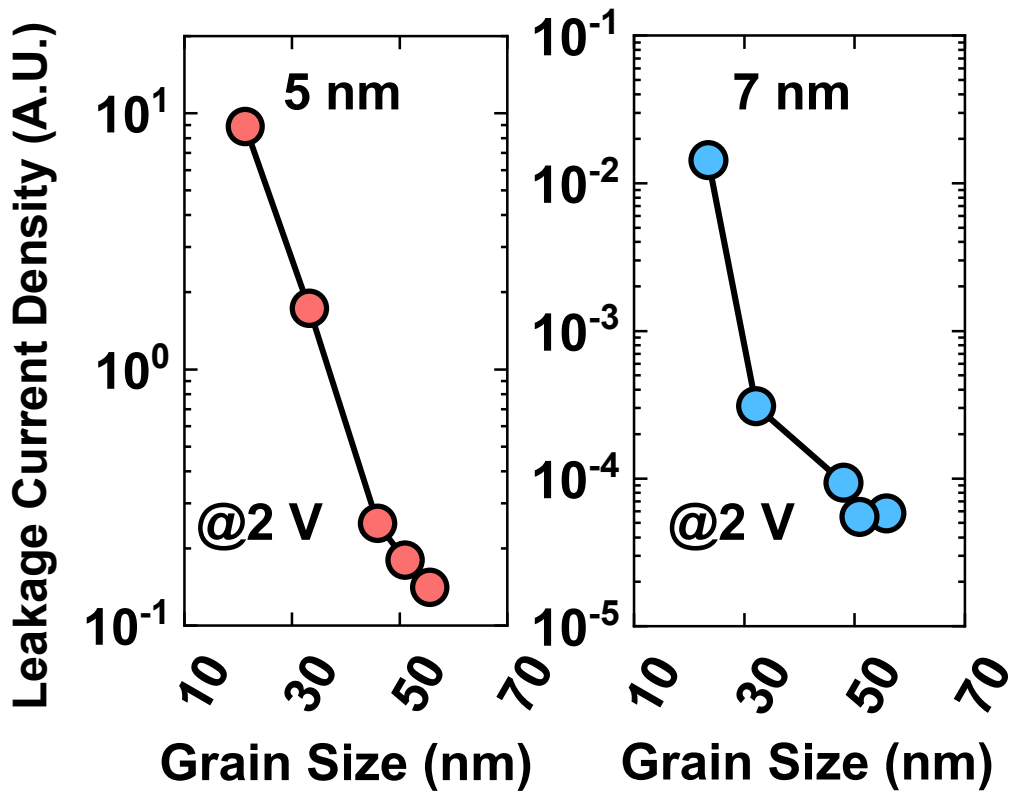

**Fig. 11.** Mean grain size dependence of leakage current density of HZO with different $t$ measured at 2 V, showing exponential increase in leakage current as grain size decreases.

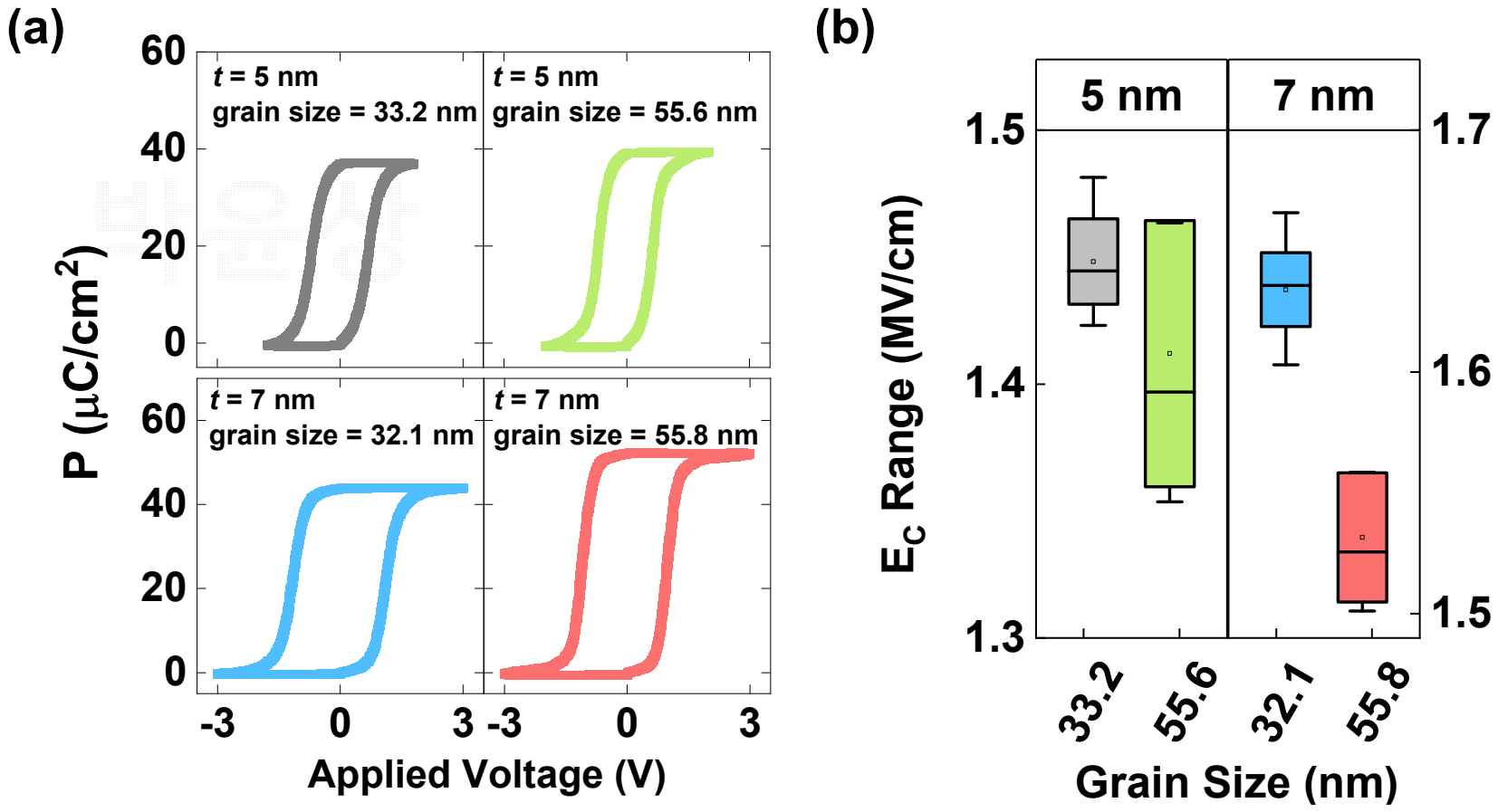

**Fig. 12.** Ferroelectric properties of HZO thin films in MFM stack with different $T_{dep}$ and $t$. **(a)** P-V hysteresis curves. **(b)** Boxplots for $E_c$ of HZO thin films with different $T_{dep}$ and $t$. Note significant reduction in σ with grain size reduction.